\documentclass[12pt]{article}
\usepackage{amsmath,amssymb,amsbsy}
\usepackage{graphicx}
\usepackage[T1]{fontenc}
\def\frame#1{\ifmmode\dframe{#1}\else\leavevmode\lower 2.4 pt
    \hbox{\vrule\unskip\vbox{\hrule\kern 1.5 pt\hbox{\kern
    1.5 pt{#1}\kern 0.5 pt}\kern 2 pt\hrule}\unskip\vrule}\fi}

\def\dframe#1{\hbox{\vrule\unskip$\vcenter{\hrule\kern 3 pt\hbox
    {\kern 3 pt$\displaystyle{#1}$\kern3pt}\kern 3 pt\hrule}$\vrule}}

\def\pni{\par \noindent}
\def\vsh{\vskip 0.1truecm}

\def\vsp{\vsh \pni}

\def\q{\quad}  \def\qq{\qquad}

\def\ds{{\displaystyle}}
\def \rec#1{{1\over{#1}}}
\def\r{\right} \def\l{\left}
\def\rt{\right} \def\lt{\left}
\def\RR{\vbox {\hbox to 8.9pt {I\hskip-2.1pt R\hfil}}}

\def\CC{{\rm C\hskip-4.8pt \vrule height 6pt width 12000sp\hskip 5pt}}
\def\e{{\rm e}}
\def\ds{\displaystyle}
\def\eg{{\it e.g.}\ }
\def\ie{{\it i.e.}\ }
\def\d{\partial}
\def\dx{\partial x}    \def\dt{\partial t}

\def\RR{\vbox {\hbox to 8.9pt {I\hskip-2.1pt R\hfil}}}
\def\CC{{\rm C\hskip-4.8pt \vrule height 6pt width 12000sp\hskip 5pt}}

\def\barr{\widetilde}

\def\Js{{\widetilde J(s)}}
\def\Gs{{\widetilde G(s)}}

\def\L{{\cal L}} 

\def\RR{\vbox {\hbox to 8.9pt {I\hskip-2.1pt R\hfil}}}
\def\CC{{\rm C\hskip-4.8pt \vrule height 6pt width 12000sp\hskip 5pt}}

\def\Gc{{\cal {G}}_c}	\def\Gcs{\barr{\Gc}} 
\def\Gs{{\cal {G}}_s}	\def\Gss{\barr{\Gs}} 

\def\arg{ x^2/ (4\,a\, t)}

\def\frame#1{\ifmmode\dframe{#1}\else\leavevmode\lower 2.4 pt
    \hbox{\vrule\unskip\vbox{\hrule\kern 1.5 pt\hbox{\kern
    1.5 pt{#1}\kern 0.5 pt}\kern 2 pt\hrule}\unskip\vrule}\fi}

\def\dframe#1{\hbox{\vrule\unskip$\vcenter{\hrule\kern 3 pt\hbox
    {\kern 3 pt$\displaystyle{#1}$\kern3pt}\kern 3 pt\hrule}$\vrule}}

\begin{document}

\title{
{\bf Fractional Calculus \\  in Wave Propagation Problems}}
\author{{\bf Francesco MAINARDI}
 \vspace{0.25truecm}
 \\ Department of Physics, University, and INFN, \\  
Via Irnerio 46, I-40126 Bologna, Italy\\  
E-mail: {\tt francesco.mainardi@unibo.it}\\
URL: {\tt http://www.fracalmo.org}}
\date{}
 \maketitle
\noindent  
{This paper, presented as an invited lecture at Fest-Kolloquium for celebrating 
the 80-th anniversary of Prof. Dr. Rudolf Gorenflo, held at the Free University of Berlin on 24 June 2011, 
has been published in {Forum der Berliner Mathematischer Gesellschaft}, Vol 19, pp. 20--52 (2011).}
 
\vskip 0.25truecm
 
\hfill{{\it Dedicated to Professor Rudolf Gorenflo}}

\hfill{{\it on the occasion of his 80th anniversary}}
\section*{Abstract}
 Fractional calculus, in allowing integrals and derivatives of any
positive order (the term "fractional" kept only for historical reasons), can be
considered a branch of mathematical physics which mainly deals with integro-differential
equations, where integrals are of convolution form with weakly
singular kernels of power law type. In recent decades fractional calculus has won
more and more interest in applications in several fields of applied sciences. In this
lecture we devote our attention to wave propagation problems in linear
viscoelastic media.
Our purpose is to outline the role of fractional calculus in providing
simplest evolution processes which are intermediate between diffusion and
wave propagation. The present treatment mainly reflects the research activity and
style of the author in the related scientific areas during the last decades.
\newpage
\section*{Sommario}
Il calcolo frazionario tratta 
di integrali e derivate di  ordine positivo qualsiasi.
Si noti che il termine "frazionario" \`e mantenuto solo per ragioni storiche.
Esso pu\`o essere considerato una branca della Fisica Matematica che principalmente
analizza  equa\-zioni integro-differenziali in cui gli integrali sono di tipo convolutivo 
con nuclei debolmente singolari a  legge di potenza.
Recentemente  il calcolo frazionario ha acquistato un interesse crescente per le applicazioni che trova in vari 
campi delle scienze applicate.
In questa lezione noi rivolgiamo l'attenzione a problemi di propaga\-zione ondosa
in mezzi lineari viscoelastici.
Il nostro proposito \`e di sottolineare il ruolo del calcolo frazionario nel fornire
semplici processi di evolu\-zione che sono intermedi tra la diffusione e la propagazione di onde.
Il trattamento presente  si basa principalmente sull'attivit\`a di ricerca dell'autore.
\section*{Acknowledgements}
\vsp
The author  appreciates the invitation 
of the {Berlin Mathematical Society} 
and of the {Department of Mathematics and Informatics of the
Free University of Berlin}  that have provided him with the opportunity to honour 
{Professor Rudolf Gorenflo} to whom he is grateful for long-lasting collaboration.
\vsp
This lecture is based on the  author's recent book  
{"Fractional Calculus and Waves in Linear Viscoelasticity"}, Imperial
College Press, London (2010), pp. 340, ISBN 978-1-84816-329-4.

\pagebreak 
\section{Introduction}
In this lecture  we devote our attention to the applications of fractional calculus 
 in providing	the simplest evolution processes
which are intermediate between diffusion and wave propagation.
As an example we consider special linear viscoelastic media that, by exhibiting power law creep,   
turn out to be intermediate models between viscous fluids (diffusion)  
and elastic solids (waves).
\vsp
We start to consider  the family of
evolution equations  obtained from the standard
diffusion equation (or the D'Alembert  wave equation)
by replacing  the	first-order (or the second-order) time derivative
 by a fractional  derivative ({\bf in the Caputo sense})
of order  $\beta$ with $0 <\beta \le 2\,, $  namely
  $$
  \frame{
  {\d^{\beta} w\over \dt^{\beta} } =
  a\,{\d^2 w \over \dx^2}  \,,	\qq  a>0\,, \q 0<\beta \le 2\,,}
\eqno(1.1)$$
where  $x\in S \subset \RR $, $t\in \RR^+$ denote the space and time
variables, respectively.
\vsp
In Eq. (1.1) $w(x,t)$ represents the response field variable,  $a$ is a positive
constant of dimension $L^2 \, T^{-\beta}\,.$
For essentials on the Caputo fractional derivative we refer the reader to
the appendix; more details can be found
in the literature on fractional calculus, see e.g. Gorenflo \& Mainardi (1997),
Podlubny (1999), Kilbas, Srivastave \& Trujillo (2006). 
\vsp
It is necessary  to keep distinct  the cases
 $$0<\beta \le 1\,, \q \hbox{and} \q  1<\beta \le 2\,,$$ 
 by recalling 
 \pni for $0<\beta <1$ :
$$ \frame{ 
{\d^{\beta} w\over \dt^{\beta} } :=  \
  {\ds \rec{\Gamma(1-\beta )}} \,
  {\ds \int_0^t}   \left[{\ds{\d \over \d \tau}} \, w(x,\tau)\right]\,
  {\ds{d\tau \over (t-\tau)^{\beta}}}  \,,} 
\eqno (1.2)$$ 
\pni for $1 < \beta<2$ :
$$\frame{
{\d^{\beta} w\over \dt^{\beta} } :=  
 {\ds\rec{\Gamma(2-\beta )}} \,
  {\ds \int_0^t}   \left[{\ds{\d^2 \over \d \tau^2}} \, w(x,\tau)\right]\,
  {\ds{d\tau \over (t-\tau)^{\beta -1}}}  \,,}
  \eqno(1.3)$$
where $\Gamma$ denotes the Gamma function. 
\vsp
We also outline that the expression in the R.H.S. of (1.2) as $\beta \to 1^-$ 
reduces to the standard derivative of order 1,
whereas  the  corresponding expression in  (1.3) as $\beta \to 2^-$ reduces to the standard derivatives of order 2.
\vsp
It should be noted that 
for $0<\beta<1$ and $1<\beta <2$,
in view of (1.2) and (1.3),
Eq (1.1)  turns out to be an 
integro-differential equation with a weakly singular kernel.
The singularities can be removed by a suitable fractional  integration in time, 
taking into account the necessary initial   conditions at $t=0^+$. 
Consequently we get  the integro-differential equations
\pni for $0<\beta \le 1$ :
$$  w(x,t) =
w(x,0^+)
 + {\ds{a\over \Gamma(\beta )}}\,
  {\ds \int_0^t}
 \left({\ds{\d^2 w\over \dx^2}}\right) \,  (t-\tau)^{\beta -1}\, d\tau\,;
   \eqno(1.4)$$
\pni for $1<\beta \le 2$ :
$$ 
\begin{array}{ll}
w(x,t) =
 & w(x,0^+)     + t \, w_t(x,0^+) \\
 & + {\ds{a\over \Gamma(\beta )}}\,
  {\ds\int_0^t }
 \left({\ds{\d^2 w\over \dx^2}}\right)  (t-\tau)^{\beta -1}\,	\, d\tau\,.
 \end{array}
    \eqno(1.5)$$

\vsp
There is huge literature concerning evolution equations of
the  types discussed above,
both with and without reference to the fractional calculus.
We  quote a  number of  references in the last decades of the
past century, that have 
mostly attracted our attention, e.g.,
Caputo (1969, 1996),
Meshkov \& Rossikhin (1970),
Pipkin (1972-1986),
Gonovskii \& Rossikhin (1973),
Buchen \&  Mainardi (1975),
Kreis \& Pipkin (1986), Nigmatullin (1986),
Fujita (1989a,1989b), Schneider \& Wyss (1989),
Kochubei (1990),
Giona \&  Roman (1992), 
Pr\"usse (1993),
Metzler {\it et al.} (1994),
Engler (1997),
Rossikhin \&  Shitikova (1997, 2007, 2010).
This talk is a brief survey of my work carried out  since 1993 when I started 
to re-consider wave propagation problems  by using the methods of the fractional calculus
\footnote{
I became aware of fractional calculus since the late 1960's as a PhD student of Prof. Michele Caputo:
this led to two  papers in the framework of linear viscoelasticity, see Caputo \& Mainardi (1971a), (1971b).
However, Mmy first approach to  Fractional Calculus was  a source of  disappointment
due to the bad reaction of the great majority of the scientific community  in that time. It was only with the
advent of fractals fashion  that more and more scientists start to consider that mathematical models based
 on  fractional calculus could be successfully adopted to explain certain physical phenomena like anomalous 
 relaxation, anomalous diffusion, etc.}     
\newpage
\vsp
{\bf The plan of the lecture is as follows}.
\vsp 
In {\bf Section 2} we derive the general evolution equation
governing the  propagation of uniaxial stress waves,
in the framework of the  dynamical theory
of linear viscoelasticity.
For a power-law solid exhibiting a creep law proportional
to $t^\gamma$ ($0 < \gamma <1$)
the evolution equation is shown to be of type  (1.1)
with 
$$	1< \beta := 2 -\gamma <2\,.\eqno (1.6)$$
\vsp
In {\bf Section 3} we  review the analysis of the fractional
evolution equation (1.1)
in the	general case $0 <\beta	< 2\,. $
\vsp
We first analyze
the two basic boundary-value problems
referred to as
the {\bf Cauchy} problem  and the {\bf Signalling} problem,
by   the technique of  the Laplace transforms
 and we derive	the transformed expressions
of the respective fundamental solutions ({\bf the Green functions}).
\vsp
Then, we  carry out the inversion of the relevant  transforms
and we outline a {\bf reciprocity relation} between the Green functions
in the space-time domain. 
\vsp
 In view of this relation
the Green functions can be expressed
in terms  of two interrelated {\bf auxiliary functions}  in
the similarity variable $r = |x|/(\sqrt{a}t^\nu  )\,,$
where $\nu  =\beta /2\,. $
These auxiliary
functions can be analytically continued in the whole complex plane
as entire functions of {\bf Wright} type.
\vsp
In {\bf Section 4} we outline the {\bf scaling properties} of the fundamental
solutions and we
exhibit their evolution
for some values of the order $\beta \,. $
For $1<\beta < 2$
the behaviour of the fundamental solutions turns out to be intermediate
between diffusion (found for a viscous fluid)  and
wave-propagation  (found for an elastic solid), thus
justifying the attribute of {\bf fractional diffusive waves}.
\vsp
In {\bf Section 5}  the fundamental solutions  are
interpreted as probability density functions related
to   {\bf L\'evy stable processes}
with index of stability  depending on $\beta \,. $
\vsp
In {\bf Appendix A} we recall the essentials of the {\bf time--fractional differentiation} 
whereas in {\bf Appendix B} we exhibit some graphical representations of the relevant Wright function. 
\section{Linear viscoelastic waves and \\ the
 fractional diffusion-wave equation}
According to the elementary one-dimensional theory of 
{\bf linear viscoelasticity}, the medium  is assumed to be homogeneous
(of density $\rho $), semi-infinite or infinite in extent
($0 \le x<+\infty $ or	 $-\infty <x < +\infty$) and
undisturbed for $t<0\,. $
\vsp
The basic equations are known to be,
see \eg Hunter (1960), Caputo \& Mainardi (1971b), Pipkin (1972-1986),
Christensen (1972-1982), Chin (1980), Graffi (1982),
$$ \sigma _x(x,t)  = \rho \, u_{tt}(x,t)\,, \eqno(2.1)$$
$$  \varepsilon (x,t) = u_x(x,t)\,, \eqno(2.2)$$
$$  \varepsilon (x,t) = [J_0  + \dot J(t) * \,] \,\sigma (x,t)
    \,. \eqno(2.3)$$
The following notations have been used:
$\sigma $ for the stress, $\varepsilon  $	for  the strain,
$J(t) $ for  the creep compliance (the strain response to a
unit step input of stress); the constant
$J_0 :=  J(0^+) \ge 0$ denotes
 the instantaneous (or glass) compliance.
\vsp
The evolution equation for the {\bf response variable}
 $w(x,t)$ (chosen among
the field variables:
the displacement $u$, the stress $\sigma $, the strain $\varepsilon $
or the particle velocity $v=u_t$) can be derived through the
application of the {\bf Laplace transform} to the basic equations.
 $$ {\cal{L}}\, \l\{  f(t);s \r\}	:= \int_0^\infty \!\!
  \e^{-st}\, f(t)\, dt = \widetilde f(s) \div f(t)\,, \quad s \in \CC\,.$$
We first obtain in the transform domain,
the  second order differential equation 
$$\frame{
\l[ {d^2\over dx^2 }- \mu ^2(s) \r]\, \widetilde w(x,s)=0\,,} \eqno(2.4)
$$ in which
$$\mu (s) := s\, \l[\rho \, s \Js \r]^{1/2}\,,	\eqno(2.5)  $$
is real and positive for $s$ real and positive.
As a matter of fact, $\mu (s)$ turns out to be an analytic
function of $s$ over the entire $s$-plane cut along
the negative real axis; the cut  can be limited or unlimited
in accordance with the particular visco\-elastic model assumed.
\vsp
Wave-like or diffusion-like  character of the evolution equation
 can be drawn from (2.5) by taking into
account the asymptotic representation of the creep compliance
for short times,
  $$\frame{
J(t) = J_0  +O(t^\gamma)\,, \q {\rm as}\; t \to 0^+\, ,}\eqno(2.6)$$
 with $J_0 \ge 0\,,$ and $0<\gamma  \le 1  \,.$
\vsp
If $J_0 >0$ then
$$ \lim_{s\to \infty} {\mu (s)\over s} = \sqrt{\rho J_0}
   := {1 \over c}\,,   \eqno(2.7)$$
 we have a {\bf wave like behaviour} with $c$ as the wave-front velocity;
otherwise ($J_0 =0)$ we have a {\bf diffusion like behaviour}.
\vsp
In the	case $J_0>0$ the wave-like
evolution equation for $w(x,t)$
can be derived by inverting (2.4-5), using (2.6-7)
and introducing the non-dimensional rate of creep
 $$  \psi(t) :=  {1\over J_0}\,  {dJ(t)\over dt} \ge 0\,,
   \q t>0\,. \eqno(2.8)$$
We get
$$  \mu ^2(s) := s^2 [\rho \, s\Js]
= [1 + \widetilde \psi(s)]\,  {s^2\over c^2}  \,, \eqno(2.9)$$
so that the evolution equation turns out as
$$\frame{
 \left\{ 1 + \psi(t) \, *\, \right\}  \, {\d^2 w\over \dt^2}
       = c^2\,{\d^2 w\over \dx^2}
    \,.} \eqno(2.10)$$
This is  a generalization of  D'Alembert  wave equation
in that it is an integro-differential equation
where the   convolution integral can be interpreted as a
perturbation term.
This case has been investigated by Buchen and  Mainardi (1975)
and by Mainardi and Turchetti (1975), who have provided
wave-front expansions for the solutions.
\vsp
In the	case $J_0 =0$ we can re-write (2.6)  as
   $$  
   \frame{J(t) = {1\over \rho \,a} \, {t^\gamma   \over \Gamma(\gamma  +1)}
  +  o\, (t^\gamma  ) \,,\q {\rm as}\; t \to 0^+\,,}	  \eqno(2.11)$$
where, for  convenience, we have  introduced
 the positive constant $a$ (with dimension
$L^{2}\, T^{\gamma  -2}$)
 and  the Gamma function $ \Gamma(\gamma  +1)\,. $
Then we can introduce
the non-dimensional function  $\phi(t)$  whose
Laplace transform is such that
$$ \mu ^2(s) := s^2 \,[\rho \, s\Js]
  = [1 +\widetilde\phi(s)]\,{s^{2-\gamma  } \over a}
  \,. \eqno(2.12)$$
Using (2.12), the Laplace inversion of (2.4-5)	 yields
$$\frame{
 \left[ 1 + \phi(t) \, *\, \right]  \, {\d^\beta w \over \dt^\beta }
       = a\,{\d^2 w\over \dx^2}
    \,, \q \beta  = 2 - \gamma\,,}
	\eqno(2.13)$$
so that, being $0<\gamma \le 1$, we have  $ 1\le  \beta  <2$.
\vsp
When
the creep compliance satisfies the simple power-law
$$ \frame{
J(t) =    {1\over \rho \,a} \, {t^\gamma	 \over \Gamma(\gamma  +1)}
\,, \q	0<\gamma \le 1
\,,\q t > 0\,,}	  \eqno(2.14)$$
we obtain $\mu ^2(s) = s^{2-\gamma}/a$ so $\phi(t) \equiv 0\,. $
As a consequence  the evolution equation (2.13) simply reduces to  Eq. (1.1).
As pointed out by  Caputo and Mainardi  (1971b),	
 the creep law (2.14)
is provided by	viscoelastic  models whose   stress-strain relation
(2.3)  can be simply expressed	by a fractional derivative
of order $\gamma  \,. $ In the present notation this stress-strain
relation reads
$$ \frame{
\sigma  = { \rho \,a}\,{d^\gamma	  \over dt^\gamma  }\,\varepsilon
 \,, \q 0<\gamma  \le 1\,.} \eqno(2.15)$$
For $\gamma	=1$ the Newton law for a viscous fluid is recovered
from (2.15) where $a$ now represents the kinematic  viscosity;
in this case, since $\beta  =1$ in (1.2),
the {\bf classical diffusion equation}  holds for $w(x,t)\,.$
\vsp
In the limiting case $\gamma=0$ we obtain from (2.14)
$$J(t) = J_0 = 1/(\rho \,  a)$$ 
so we recover from (2.10) and (2.12)
the {\bf classical D'Alembert wave equation} ($\beta =2$) with
wave front velocity $c= \sqrt{a}\,. $
\vsp
When $ 0<\gamma  <1$ we just obtain the evolution  equation (1.1)
with $1<\beta <2$ In this case, as we have previously pointed out,
being  intermediate between
the heat equation   and the wave equation, 
Eq. (1.1) is referred to as  the {\bf fractional
diffusion-wave equation}, and its solutions  can be interpreted as 
fractional diffusive waves, see Mainardi (1995), Mainardi \& Paradisi (2001).
\vsp
We point out that the viscoelastic models based on (2.14) or (2.15)
with $0<\gamma  <1$
and henceforth governed by the fractional diffusion-wave equation,
are of great interest in material sciences and seismology.
In fact, as shown by Caputo \& Mainardi (1971b) and then 
by Caputo (1973, 1976, 1979), 
these models
exhibit an internal friction 
independent on frequency according to  the law
$$ 
\frame{
Q^{-1} = {\rm tan}\, \left(\gamma  \,\pi\over 2\right) \,
\Longleftrightarrow \, \gamma  = {2\over \pi}\, {\rm arctan}\,
   \left ( Q^{-1} \right) \,.} \eqno(2.16)$$
The independence of the $Q$ from the frequency is in fact
experimentally verified  in pulse propagation phenomena
for many materials, see Kolsky (1956) including  those of seismological interest,
see Kjiartansoon (1979), Strick (1970, 1982), Strick and Mainardi (1982).
\vsp
From (2.16) we note that  $Q$ is also independent
on the material constants $\rho $ and $a$
which, however, play a role in the phenomenon of wave dispersion.
\vsp
The limiting cases of absence of energy dissipation (the elastic energy
is fully stored) and  of absence of energy storage (the elastic
energy is fully dissipated) are recovered from (2.16)
for $\gamma	=0\, $ (perfectly elastic solid)
and $\gamma	=1\,$ (perfectly viscous fluid), respectively.
\vsp
To obtain   values of seismological interest for the dissipation
 ($  Q \approx 1000$)	we need to choose the parameter $\gamma  $
sufficiently  close to zero, which corresponds
to a {\bf nearly elastic} material; from (2.16) we obtain the
approximate relations between $\gamma  $ and $Q\,, $ namely
$$\frame{  
\gamma	\approx \left( {2\over \pi\, Q}\right)
     \approx 0.64  Q^{-1}  
  \Longleftrightarrow 
      Q^{-1}  \approx {\pi\over 2} \gamma
     \approx 1.57  \gamma.} \eqno(2.17)$$
\section{Derivation of the fundamental solutions} 
In order to  guarantee the existence and the uniqueness of the
solution, we must equip (1.1) with suitable data on the
boundary of the space-time domain.
\vsp
The  basic boundary-value problems
for  diffusion
are the so-called {\bf Cauchy} and {\bf Signalling} problems.
\vsp
In the	{\bf Cauchy problem}, which concerns  the space-time domain
$-\infty <x< + \infty\,, $ $\, t \ge 0\,, $
the data are assigned at $t=0^+$ on
the  whole space axis (initial data).
\vsp
In the {\bf Signalling problem}, which concerns  the space-time domain
$x\ge 0\,, $ $\, t \ge 0\,, $
the data are assigned both at $t=0^+$ on the semi-infinite
space axis $ x >0 $ (initial data) and at $x=0^+$ on the semi-infinite
time axis  $ t>0$ (boundary data); here, as mostly usual,
the initial data are assumed to  vanish.
\vsp
Denoting by
$f(x)\,,\, x\in \RR\,$ and $\,h(t)\,,\, t\in \RR^+\,$
sufficiently
well-behaved functions,    the basic problems are thus formulated as
following, assuming $0<\beta \le 1\,$:
\vsh\pni
a) {\bf  Cauchy problem}
$$\begin{cases}  
w(x,0^+) =f(x) \,,	\q -\infty <x < +\infty\,;\\
w(\mp \infty,t) = 0\,,\q \, t>0\,;
\end{cases}
  \eqno(3.1a) $$
\pni
b) {\bf Signalling problem}
$$  \begin{cases}
w(x, 0^+)  =  0 \,, \q  x>0\,;\\
    w(0^+,t ) =h(t) \,,\; w(+\infty,t) =0 \,, \q   t >0
\,. 
\end{cases}
\eqno(3.1b) $$
If $1 <\beta < 2\,, $ we must add in (3.1a) and (3.1b)
the initial values of the first time derivative of the field variable,
$w_t(x,0^+)\,,$ since in this case
the  fractional derivative is expressed in terms
of the second order  time derivative.
To ensure the continuous dependence  of our solution
with respect to the parameter $\beta  $
also in the transition from $\beta  =1^-$ to  $\beta =1^+\,,$
we agree  to  assume 
$$w_t(x,0^+) = 0\,, \q \hbox{for} \q 1<\beta \le 2\,,$$
as it turns out from the integral forms (1.4)-(1.5).
\vsp
In view of our subsequent analysis we find it convenient to set
$$  \nu  :={\beta / 2}\,,
   \q {\hbox{so}}\q
   \begin{cases}
    0<\nu  \le 1/2 \,, \; if \; 0<\beta\le 1\,,\\
	1/2 <\nu \le 1\,, \; if \; 1<\beta \le 2\,,
	\end{cases} 
	 \eqno(3.2)
	$$
and from now on to add the parameter  $\nu $ to the
independent space-time variables $x\,,\,t$ in the solutions,
writing $w = w(x,t;\nu )\,. $
\vsp
For the {\bf Cauchy} and {\bf Signalling}  problems we
introduce
the so-called Green functions $\Gc (x,t;\nu )$ and $\Gs(x,t;\nu )$,
which represent the respective fundamental solutions,
obtained when $f(x) = \delta (x)$ and $h(t) = \delta (t)\,. $
As a consequence, the solutions of the two basic problems
are obtained by a space or time convolution  according to
$$ 
w(x,t;\nu)
= \int_{-\infty}^{+\infty} \Gc(x-\xi ,t;\nu  ) \, f(\xi)
\, d\xi     \,,  \eqno(3.3a)$$
$$ w(x,t;\nu ) = \int_{0^-}^{t^+} \Gs(x,t-\tau;\nu  ) \, h(\tau ) \,
d\tau	  \,.  \eqno(3.3b)$$
It should be noted that in (3.3a) $\Gc(x ,t ;\nu) =
  \Gc(|x|,t ;\nu)$ since the Green function of the Cauchy problem turns
out to be an even function of $x$.
According  to a usual convention,
in (3.3b) the limits of integration are extended to take into account
for the possibility of impulse functions centred at the extremes.
For the  standard {\bf diffusion} equation ($\nu   =1/2$)
it is well known that
$$   
\Gc(x,t;1/2) := \Gc^d (x,t)
 = {t^{-1/2}\over 2\sqrt{\pi a}}\,\e^{-\ds \arg}\,,
 \eqno(3.4a) $$
$$ \Gs(x,t;1/2) := \Gs^d (x,t)
 = {x\, t^{-3/2}\over 2\sqrt{\pi a }} \, \e^{-\ds \arg}\,.
 \eqno(3.4b)$$
In the limiting case $\nu  =1$ we recover  the standard {\bf wave equation},
for which, putting $c = \sqrt{a }\,, $
$$   
\Gc(x,t;1) := \Gc^w (x,t)
 ={1\over 2} \left[  \delta (x-ct) + \delta (x+ct)\right] \,,
 \eqno(3.5a) $$
$$\Gs(x,t;1) := \Gs^w (x,t)
 = \delta (t-x/c)  \,. \eqno(3.5b) $$
In the general case $0<\nu  <1$ the two Green functions will be
determined by using the technique of the Laplace transform.
\vsp
 For the {\bf Cauchy problem} (3.1a)   with $f(x) = \delta (x)$
the application of the Laplace
transform to  Eq. (1.1) with
$w(x,t)=\Gc(x,t;\nu )\, $
leads  to the non homogeneous differential equation
satisfied by the image of the Green function,
$\Gcs(x,s;\nu  )\,, $
$$
 a\,{d^2 \Gcs  \over dx^2}-  s^{2\nu }	\,\Gcs=
   - \,\delta (x)\,s^{2\nu  -1}\,, \; -\infty <x<+\infty
    \,.       \eqno(3.6)$$
Because of the singular term $\delta (x)$ we have to consider
the above equation separately in the two intervals $x<0$ and $x >0$,
imposing the   boundary conditions  at $x=\mp \infty\,, $
$$\Gc({\mp \infty, t;\nu }) =0\,, $$
and the necessary matching conditions
at $x= 0^{\pm}$.
\vsp
We obtain
$$ \frame{
 \Gcs(x,s;\nu ) =
    \frac{\e^{\ds  -(|x|/\sqrt{a })\,s^\nu}}{2\sqrt{a }\, s^{1-\nu }} \,, \; -\infty <x<+\infty
\,.} \eqno(3.7)$$
 \vsp
 For the {\bf Signalling problem} (3.1b) with $h(t) = \delta (t)$ the
application of the Laplace transform to Eq. (1.1) with
$w(x,t)=\Gs(x,t;\nu )\, $
 leads to   the  homogeneous differential equation
$$   a \,{d^2  \Gss \over dx^2}
   -s^{2\nu  } \,\Gss	= 0 \,, \q x\ge 0  \,.	     \eqno(3.8)$$
Imposing the boundary conditions at $x=0\,, $
$\Gs(0^+,t;\nu )=h(t) =  \delta (t)\,, $
and at $x=+\infty\,, $
$\Gs(+\infty,t;\nu )=0\,, $
we obtain
$$  \frame{\Gss(x,s;\nu  )  =
 \e^{\ds  -(x/\sqrt{a })s^{\nu	}}\,, \q x\ge 0\,.} \eqno(3.9)$$
\vsp
From (3.7) and (3.9) we recognize
for the  original Green functions
 the following {\bf reciprocity relation}
$$\frame{ 2\nu  \, x\,  \Gc(x,t;\nu  )  = t\, \Gs(x,t;\nu  )
  \,,\q x > 0\,,\q t > 0 \,.}\eqno(3.10) $$
  \vsp
This  relation	can be easily verified in the case of standard
diffusion ($\nu  =1/2$),
where the explicit expressions
(3.4a)-(3.4b) of the Green	functions leads to the identity for $x>0,\, t>0$,
    $$ 
	\begin{array}{ll}
	x\, \Gc^d (x,t) & = t\, \Gs^d (x,t) = {1\over 2\sqrt{\pi}}
   \, {x \over \sqrt{a \,t}}\,	\e^{-\ds\arg}\\
   &= F^d(r) =  {r\over 2}\,M^d(r)\,,
   \end{array}
 \eqno(3.11)$$
where
 $$ M^d(r)= \rec{\sqrt{\pi}} \,\e^{-\ds  r^2/4}\,, \q
 r={x \over \sqrt{a}\, t^{1/2}} > 0\,.
\eqno(3.12)$$
The variable $r$  is the well-known {\bf similarity variable} whereas
the two functions $F^d(r)$ and $M^d(r)$ can be considered
the {\bf auxiliary functions}
for the diffusion equation because each of them
provides the fundamental solutions through (3.11).
\vsp
We note that $M^d(r)$ satisfies the normalization condition
$$ \int_0^{\infty} \! M^d(r)\, dr =1\,.$$
\vsp
In terms of the auxiliary functions the reciprocity relation (3.10) reads
(for $x>0, \, t>0$)
$$ \frame{2\nu  \, x\,  \Gc(x,t;\nu  )  = t\, \Gs(x,t;\nu  )
    = F_\nu(r) = \nu  r\, M_\nu(r)
  \,.}\eqno(3.13) $$
where
$$r={x/(\sqrt{a}\, t^{\nu  })} >0\,\eqno(3.14)$$
is the {\bf similarity variable} and
$$\frame{
\begin{cases}
 F_\nu(r) := 
 {\ds {1\over 2\pi i}\,\int_{Br}   \!\!
 \e^{\ds  \sigma -r\sigma ^\nu } \,
   {d\sigma} }\,, \\ \\
 M_\nu(r) :=
 {\ds {1\over 2\pi i}\,\int_{Br}   \!\!
 \e^{\ds  \sigma -r\sigma ^\nu } \,
   {d\sigma \over \sigma^{1-\nu }}} 
  \end{cases} \; 0<\nu<1\,,}
    \eqno(3.15)$$
are the two {\bf auxiliary functions}.
\vsp
In (3.15)  $Br$ denotes the {\bf Bromwich path}
that can be deformed into the {\bf Hankel path} $Ha$. 
\vsp
Then, the integral and series representations
of $F_\nu(z)$ and  $M_\nu(z)$, valid on all of $\CC\,,$
with $0<\nu  <1$
turn out to be
$$ \frame{
\begin{array}{ll}
F_\nu(z) =  
 & {\ds  {1\over 2\pi i}\,\int_{Ha}	\!\!
 \e^{\ds  \, \sigma -z\sigma ^\nu } \,
   d\sigma}  =
{\ds \sum_{n=1}^{\infty}{(-z)^n\over n!\, \Gamma(-\nu  n)}}\\
  & = -{\ds{1\over \pi}\,\sum_{n=1}^{\infty}{(-z)^n\over n!}\,
    \Gamma(\nu	n +1 )\, \sin(\pi \nu	n)}\,,
 \end{array}}
   \eqno(3.16)$$
 and
$$
\frame{
\begin{array}{ll}
 \!\!M_\nu(z) = 
 &
 {\ds	  {1\over 2\pi i}\,\int_{Ha}   \!\!
 \e^{\ds  \,\sigma -z\sigma ^\nu } \,
   {d\sigma\over \sigma ^{1-\nu  }}}\!=\!
{\ds \sum_{n=0}^{\infty}
 {(-z)^n \over n!\, \Gamma[-\nu  n + (1-\nu  )]}}\\
 & =  {\ds  \rec{\pi}\, \sum_{n=1}^{\infty}\,{(-z)^{n-1} \over (n-1)!}\,
  \Gamma(\nu  n)  \,\sin (\pi\nu  n)}\,.
  \end{array}}
 	\eqno(3.17)$$
\vsp
In the theory of special functions, see Ch 18 in Vol. 3 of the handbook
of the Bateman Project, see Erd\'ely (1955),
we find an entire function, referred to as the
{\bf Wright function},	which reads (in our notation) for $z\in \CC$:
  $$ \frame{
  \begin{array}{ll}
  W_{\lambda ,\mu }(z ) & :=  {\ds  {1\over 2\pi i}\,\int_{Ha}   \!\!
 \e^{\, \ds \sigma +z\sigma ^{-\lambda }} \,
   {d\sigma \over \sigma^{\mu}}} \\
& :=   {\ds  \sum_{n=0}^{\infty}{z^n\over n!\, \Gamma(\lambda  n + \mu )}}
   \,,
   \end{array}}	
\eqno(3.18)$$
where $\lambda >-1$ and $\mu >0\,. $
From a comparison among (3.16-3.17) and (3.18) we recognize that
the auxiliary functions are  related to the Wright function according
to
$$ \frame{
\begin{cases}
{\ds F_\nu(z) =   W_{-\nu , 0}(-z) = \nu  \, z \, M_\nu(z) \,,} \\
 {\ds  M_\nu(z) =  W_{-\nu , 1-\nu }(-z)\,.}
 \end{cases}}
   \eqno(3.19)$$
\vsp
\underbar{{\bf Remark}}
\vsp
We note that in the Bateman handbook, 
presumably for a misprint, the Wright function is considered
with   $\lambda $  restricted to be non-negative.
\vsp
In his  first 1993 analysis of the time fractional diffusion equation,
 see Mainardi (1994), the present Author,
being in that time only aware of the Bateman handbook
thought to have extended the original Wright  function.
It was just {\bf Professor Stankovi{\'c}}
during the presentation of the paper
by Mainardi \& Tomirotti (1995)
in the Conference {\bf Transform Methods and Special Functions, Sofia 1994},
who informed the author that this extension for $-1<\lambda < 0$
was already made just by Wright himself in 1940 (following
his previous papers in 1930's).
\vsp
In his 1999 book, {\bf Professor Podlubny}, not aware of the remark of {\bf Professor Stankovi{\'c}}, referred
the {\bf M-Wright function} to as the {\bf Mainardi function}.  
\vsp
Although convergent in all of $\CC$, the series representations
in (3.16-17) can be used to provide a numerical evaluation of our auxiliary
functions only for relatively small  values of $r\,, $ so that
asymptotic evaluations as $r \to +\infty$ are required.
Choosing as a variable $r/\nu  $ rather than $r\,,$ the computation
by the saddle-point method for the $M$-Wright  function is
easier and yields, for $r \to +\infty$, see Mainardi \& Tomirotti (1995),
$$ M_\nu(r/\nu ) \sim
   {r^{\ds{(\nu  -1/2)/(1-\nu )}}
     \over \sqrt{2\pi\,(1-\nu )}}
   \,\e^{\ds \left[ - {\ds {1-\nu	\over \nu}}
       \,r^{\ds {1/(1-\nu )}}\right]}\,.
 \eqno(3.20)$$
We note that the saddle-point method
 for $\nu  =1/2$  provides the
exact result (3.12), \ie
 $$M_{1/2}(r) = M^d(r)= (1/\sqrt{\pi})\, {\rm exp} (-r^2/4)\,, $$
but breaks down for $\nu  \to 1^-\,. $
\vsp
The case $\nu  =1\,, $ (namely $\beta=2$)	for which
(1.1) reduces to the standard wave equation, is of course
a singular limit also for the series representation
since 
$$M_1(r)= \delta (r-1)\,.\eqno(3.21) $$ 
The exponential decay for $r \to +\infty$ ensures that
all the moments of $M_\nu (r)$  in $\RR^+$ are finite;
in particular, see Mainardi (1997), we obtain
  $$ \int_0^{+\infty} \!\!\! r^{\, n}\, M_\nu(r)\, dr
   = {\Gamma(n+1)\over \Gamma(\nu  n+1)}\,,\q n= 1\,, \, 2\, \dots
\eqno(3.22)$$
\section{The scaling properties and the evolution \\ of the fundamental solutions} 
It is known that in theoretical seismology
the delta-Dirac function is of great relevance in simulating
the pulse generated by an ideal seismic source,
concentrated in space ($\delta (x)$) or in time
($\delta (t)$).
Consequently,
the fundamental solutions of the {\bf Cauchy} and {\bf Signalling}
problems are those of greater interest because they
provide us with information on the possible evolution
of the seismic pulses  during their propagation from the seismic
source.
\vsp
Accounting for the reciprocity relation (3.13)
and  the similarity variable (3.14),
the two fundamental solutions can be written, for $x>0$
and $t>0\,, $
as
$$ \frame{ 
\Gc(x,t;\nu) = {1\over 2\, \nu\, x}\, F_\nu(r)
   =	{1\over2 \sqrt{a}\, t^\nu}\, M_\nu(r)
 \,, } \eqno(4.1a)$$
$$ \frame{
\Gs(x,t;\nu) = {1\over t}\, F_\nu(r)
 =  {\nu \, x\over \sqrt{a} \,t^{1+\nu}}\, M_\nu(r)\,.}
  \eqno(4.1b)$$
\vsp
The above equations mean that for the fundamental solution of the
 {\bf Cauchy} [{\bf Signalling}] problem the time [spatial] shape
is the same at each position [instant], the only changes
being due to space [time] - dependent changes of width and amplitude.
The maximum amplitude in time [space] varies precisely as
$1/x$ [$1/t$].
\vsp
The two fundamental solutions  exhibit scaling properties
that  make easier their plots
versus distance  (at fixed instant) and versus time (at fixed position).
In fact,
 using the well-known scaling properties of the Laplace transform
 in (3.7) and (3.9),
  we easily prove, for any  $p\,,\,q >0\,,$ that
$$\begin{cases}
 {\ds \Gc(px,qt;\nu) = {1\over q^\nu}\, \Gc (px/q^\nu ,t;\nu)\,,} \\
 {\ds \Gs(px,qt;\nu) = {1\over q}\,  \Gs (px/q^\nu ,t;\nu)\,,}
\end{cases}
  \eqno(4.2)$$
and, consequently, in plotting we can choose suitable values for the
fixed variable.
\vsp
We also note  the  exponential decay of $\Gc(x,t;\nu)$
as $x \to +\infty$ (at fixed $t$) and the algebraic decay of
 $\Gs(x,t;\nu)$
as $t \to +\infty$ (at fixed $x$), for $0<\nu <1\,. $
In fact, using (4.1a-b) with (3.17) and (3.20),
we  get for $x\to \infty$
$$  \frame{\Gc(x,t;\nu) \sim A(t)\, x^{(\nu -1/2)/(1-\nu)}
    \e^ {\ds \,-B(t) x^{1/(1-\nu)}} \,,}
\eqno(4.3a)$$
and for $t\to\infty $
 $$\frame{
 \Gs(x,t;\nu) \sim C(x)\, t^{-(1+\nu)} \,,} \eqno(4.3b)$$
where  $A(t)\,, \, B(t)$ and $C(x)$ are positive functions.
\begin{figure}[h]
\begin{center}
 \includegraphics[width=.48\textwidth]{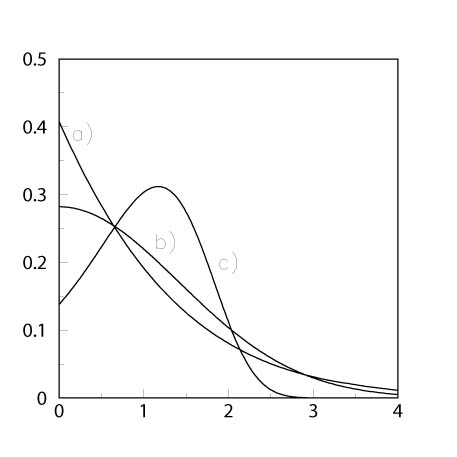}
  \end{center}
 \vskip-0.5truecm
 \caption{The Cauchy problem for the time-fractional diffusion-wave equation:
the fundamental solutions versus $|x|$  with
  a) $\nu =1/4\,, \;$	b) $\nu =1/2\,, \;$
  c) $\nu =3/4\,.$ }
 \end{figure}
 \vsp
In Figure 1, as an example
we compare versus $|x|\,, $
at fixed $t\,, $  the fundamental solutions of the {\bf Cauchy problem}
with different $\nu $ ($\nu  =1/4\,, \, 1/2\,, 3/4\,$).
We consider the range $0\le |x|\le 4$ for $t=1$, assuming $a=1$.
 \begin{figure}
\begin{center}
  \includegraphics[width=.48\textwidth]{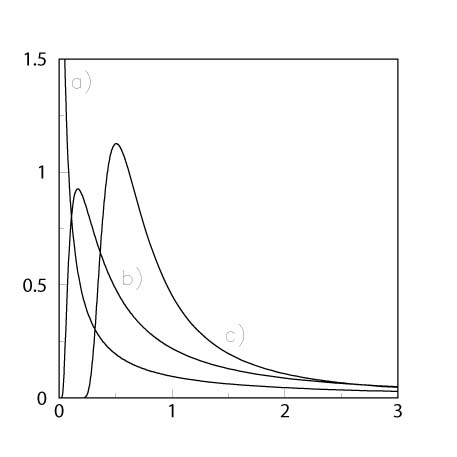}
  \end{center}
 \vskip-0.5truecm
 \caption{The Signalling problem for the time-fractional diffusion-wave equation:
the fundamental solutions versus $t$  with
  a) $\nu =1/4\,, \;$	b) $\nu =1/2\,, \;$
  c) $\nu =3/4\,.$}
  \end{figure}
\vsp
In Figure 2, as an example
we compare versus $t\,, $
at fixed $x\,, $  the fundamental solutions of the {\bf Signalling problem}
with different $\nu $ ($\nu  =1/4\,, \, 1/2\,, 3/4\,$).
We consider the range $0\le t\le 3$ for $x=1$, assuming $a=1$.
\vsp
In the limiting cases $\nu  = 0$ ($\beta=0$) and $\nu  =1$ ($\beta=2$)	we
have
$$\begin{cases}
 {\ds \Gc(x,t;0)} = {\ds{{\e}^{\,\ds -|x|}\over 2}}, & \nu=0;\\ 
 {\ds \Gc(x,t;1)} =
{\ds{\delta (x-\sqrt{a}\,t)+\delta (x+\sqrt{a}\,t)\over 2}}, & \nu=1.\\
 \end{cases}
  \eqno(4.4) $$

\vsp
In the limiting cases $\nu  = 0$ ($\beta=0$) and $\nu  =1$ ($\beta=2$)
	we have
$$\Gs(x,t;0) = \delta (t)\, , \qq
\Gs(x,t;1) = \delta (t- x/\sqrt{a})\,.
  \eqno(4.5)
$$
 \vsp
As outlined at the end of Section 2, 
in order  to ensure a sufficiently low value
(of seismological interest)
 for the constant internal friction $Q^{-1}$, we  
would inspect the evolution of the initial (seismic) pulses
$ \Gc(x,t;\nu )$ versus $x$
and  $ \Gs(x,t;\nu )$ versus $t$  when
the exponent $\gamma$ in the creep power law  (2.14)
is  close to $0$  ({\bf nearly elastic} cases).
In these limiting cases  the order $\beta$ of the fractional time derivative
 tends to 2 from below, since $\beta=2-\gamma$ from (2.13), and our exponent $\nu =\beta/2$ 
 tends to $1$ from above.
\vsp
For the analysis of the limits
 the reader is referred to Mainardi \& Tomirotti (1997)   
who have considered the evolution of the seismic pulse 
for $\nu =1-\delta $ ($\delta= \gamma/2$)  taking 
$\delta =0.001, 0.01$. 
\vsp 
 In these nearly elastic cases  the evaluation of the Green functions is indeed`a difficult task
 because the matching between their series and saddle point representations is no longer 
 achieved due to the fact that the  saddle point turns out to be wide and the consequent approximation
 becomes poor. 
\section{The fundamental solutions as   probability \\ density functions}
\def\P{{\cal{P}}}
\vsp
{\bf Cauchy Problem}:
The fundamental solution is provided by the 
{\bf Gauss} or {\bf normal}
probability density, symmetric in space.
$$\frame{  \Gc^d (x,t) =
{1\over 2\sqrt{\pi\,a\,t}} \,	\e^{-\ds\arg} =
   p_G(x; \sigma ) \,,} \eqno(5.1)$$
where
$$ p_G(x; \sigma ) :=
 {1\over \sqrt{2\pi}\,\sigma }\, \e ^{\,\ds -x^2/(2 \sigma ^2)}\,, \;
  \sigma^2 =2\, a\, t\,.
 \eqno(5.2) $$
$$ 
\begin{array}{ll}
\P_G(x;\sigma ) &:=   {\ds \int_{-\infty}^x p_G(x;\sigma )\, dx 
 = {\rec 2} \left[ 1 + {\rm erf} \left( {x\over \sqrt{2}\, \sigma }\right)\right]}\\
 & = {\ds {\rec 2} \left[ 1 + {\rm erf} \left( {x\over \sqrt{2at}}\right)\right]}
   \,.
   \end{array}
  \eqno(5.3)$$
$$\begin{array}{ll}
 {\ds \int_{-\infty}^{+\infty}x^{2n}\, p_G (x;\sigma ) \,dx}  
  &= {\ds {(2n)!\over 2^n \,n!}\, \sigma ^{2n}}\\
  &= {\ds (2n-1)!! \, \sigma ^{2n} = (2n-1)!! \, (2\,a\,t)^n}\,.
 \end{array}
\eqno(5.4)$$
\vsp
{\bf Signalling Problem}:
The fundamental solution is provided by the
{\bf L\'evy-Smirnov} probability density, unilateral in
time (a property not so well-known as that for the Cauchy problem!).
$$\frame{
 \Gs^d (x,t) = 
 {x\over 2\sqrt{\pi\, a}\,t^{3/2}}
     \e^{-\ds\arg}=
   p_{LS}(t; \mu )\,,}   \eqno(5.5)$$
$$  
p_{LS}(t ; \mu ) = 
   {\sqrt{\mu }\over \sqrt{2\pi}\, t^{3/2}}\, \e ^{\,\ds -\mu /(2t)}\,,
\;	\mu =  {x^2\over 2\,a}\,,
 \eqno(5.6)  $$
$$
\begin{array}{ll}
 {\P}_{LS}(t;\mu ) 
 &:=  {\ds \int_0^t  p_L(t ;\mu  )\, dt  =
    {\rm erfc}\, \left( \sqrt{{\mu \over 2t}}\right)}\\ 
	& =
 {{\ds \rm erfc} \,\left( {x\over 2\,\sqrt{a\,t}}\right)}
      \,.   
	  \end{array}
	  \eqno(5.7)$$
The {\bf L\'evy-Smirnov} $pdf$ has all moments  of integer order infinite,
since it decays at infinity like $t^{-3/2}$. However,
we note that
the moments of real order $\delta $ are finite only if
$ 0\le \delta  <1/2\,. $  In particular, for this $pdf$  the
mean (expectation) is infinite, but the {\bf median} is finite.
In fact, from ${\P}_{LS}(t_{med};\mu)=1/2\,, $ it turns out that
$t_{med} \approx  2 \mu \,. $ 
\vsp
The {\bf Gauss} and {\bf L\'evy--Smirnov} laws 
are special cases of the important class of $\alpha$ - {\bf stable}
probability distributions, or {\bf L\'evy stable} distributions with index
of stability  (or characteristic exponent)
$\alpha =2$ and $\alpha =1/2\,, $ respectively. 
\vsp
Another special
case  is provided for $\alpha =1$ by the 
{\bf Cauchy-Lorentz} law with $pdf$
$$p_{CL}(x;\lambda  )= \rec{\pi} \frac{\lambda} {x^2+\lambda ^2}\,,
\eqno(5.8)$$
$$
\begin{array}{ll}
\P_{CL}(x;\mu) 
 & := {\ds  \int_{-\infty}^x  p_{CL}(x;\mu )\, dx} \\
& = {\ds \rec{\pi}\, {\rm arctang} \left( \frac{x}{\lambda}  \right) +  \frac{1}{2}}\,.
\end{array}
\eqno(5.9)$$
\vsp
 The name {\bf stable} has been assigned
to these distributions	because of
the following property:
if two independent real random variables
with the same shape or {\bf type} of distribution are combined linearly and
the distribution of the resulting random variable has  the same shape,
the common distribution (or its type, more precisely) is said to be
{\bf stable}.
\vsp
More precisely,
if $Y_1$ and $Y_2$  are random variables
having such distribution, then $Y$ defined by the linear combination
$c\,Y = c_1\,Y_1 +c_2 \, Y_2$ has a similar distribution with the same
index $\alpha $ for any positive real values of the constants
$c\,,\,c_1$ and $c_2$ with
$c^\alpha = c_1^\alpha +c_2^\alpha \,. $
As a matter of fact only the range $0<\alpha \le 2$ is allowed for
the index of stability. The case $\alpha =2$ is noteworthy
since it corresponds to the {\bf normal distribution},
which is  the only stable distribution which has finite
variance, indeed finite moments of any order. In the cases $0<\alpha <2$
the corresponding $pdf$  $p_\alpha (y)$ have inverse power tails, \ie
$\int _{|y|>\lambda } p_\alpha (y)\,dy = O(\lambda ^{-\alpha })\,$
and therefore their    absolute moments
of order $\delta $ are finite if $0\le \delta <\alpha $ and
infinite if $\delta \ge \alpha \,. $
\vsp
The inspiration  for systematic research on stable distributions,
originated with Paul L\'evy,
was the desire to generalize the celebrated {\bf Central Limit Theorem}
($CLT$).
\vsp
The restrictive condition of stability enabled some authors to derive
the general form for the characteristic function ($cf$, the
Fourier transform of the  $pdf$) of
a stable distribution, see Feller (1971).
\vsp
A stable $cf$
is also {\bf infinitely divisible}, \ie for every positive integer
$n$ it can be expressed as the $n$th power of some $cf$.
Equivalently we can say that for every positive integer $n$
a stable $pdf$ can be expressed as the	$n$-fold convolution of some
$pdf\,. $
\vsp
All stable  $pdf$ are $unimodal$ and indeed {\bf bell-shaped}, \ie
their $n$-th derivative has exactly $n$ zeros,
 \vsp
The $\alpha $-{\bf stable}
distributions turn out to depend on an additional parameter
 $\theta$, the {\bf skewness}  parameter.
Denoting a stable $pdf$
by $p_\alpha (y;\theta )\,, $
we note 
$ p_\alpha (-y;-\theta ) = p_\alpha (y;\theta )\,. $
Consequently a stable $pdf$ with  $\theta =0\, $
is necessarily symmetrical.
As a matter of fact $|\theta | \le \alpha $ if $0<\alpha <1$
and  $|\theta| \le 2- \alpha \, $  if $1<\alpha <2\,,$
so the allowed region for 
$\alpha  $ and $\theta$
is the {\bf Feller-Takayasu diamond}, see Fig. 3.
\begin{figure}[h!]
\begin{center} 
\includegraphics[width=0.45\textwidth]{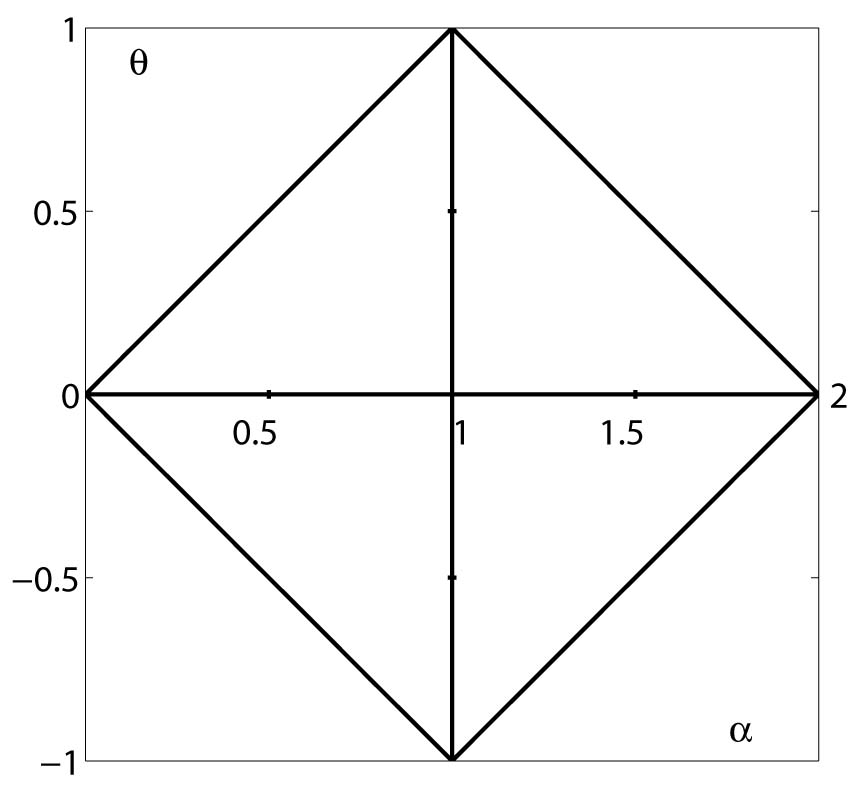}
\end{center} 
 \vskip-0.5truecm
 \caption{The Feller-Takayasu diamond}
  \end{figure}
\vsp \noindent
One recognizes that the {\bf normal distribution} is the only stable  $pdf$ independent
on $\theta $,
and that all the {\bf extremal stable distributions} with $0<\alpha  <1$
are unilateral, \ie vanishing in $\RR^\pm$ if $\theta =\pm \alpha \,. $
\vsp
In particular, the following representations by convergent power series
are valid for stable distributions with $0<\alpha <1$ (negative powers)
and $1<\alpha < 2$ (positive powers),  for $y>0\,: $
\vsp
$0<\alpha<1\,:$
$$
\frame{ 
  p_\alpha (y;\theta )= {1\over \pi\,y}  \sum_{n=1}^{\infty}
   (-y^{-\alpha})^n  {\Gamma (n\alpha +1)\over n!}
  \sin \left[{ n\pi\over 2}(\theta -\alpha)\right];}
  \eqno(5.10)$$
\vsp
$1<\alpha<2\,:$
 $$\frame{
 p_\alpha (y;\theta )= {1\over \pi\,y}  \sum_{n=1}^{\infty}
   (-y)^{n}  {\Gamma (n/\alpha +1)\over n!}
  \sin \left[{ n\pi\over 2\alpha }(\theta -\alpha)\right].}
  \eqno(5.11)$$
In the limiting case  $\alpha =2$ ($\theta=0$)
we recover the Gauss density (5.2) with $x=y$, $\sigma^2=2$.
For $\alpha =1$ and $\theta=0$ we recover the Cauchy-Lorentz density (5.8) with $x=y$, $\lambda =1$,
whereas for $\alpha =1$ and $\theta= \mp 1$ the singular densities $\delta (y \mp 1)$, 
 see Mainardi, Luchko \& Pagnini (2001).
\vsp
From Eqs. (5.10)-(5.11)  a relation    between stable $pdf$
with index $\alpha $ and $1/\alpha\,  $  can be derived.
Assuming $1/2<\alpha<1$ and $y>0\,, $  we obtain
$$ \rec{y^{\alpha +1}}\, p_{1/\alpha}(y^{-\alpha} ;\theta )
  =p_\alpha (y;\theta ^*)\,,  \;
  \theta ^*=\alpha(\theta +1)-1 \,. \eqno(5.12)$$
 A quick check shows that  $\theta^*$ falls within the prescribed range,
 $$|\theta ^*|\le\alpha \,, $$
  provided that 
  $$|\theta |\le 2-1/\alpha \,. $$
\vsp
Furthermore, we can derive a relation between extremal stable $pdf$
and  our auxiliary functions 
of Wright type.
In fact, by comparing (5.10)-(5.11) with the series representations in
(3.16)-(3.16) and using (3.19),   we obtain
\pni for  $0<\alpha <1\,,$
$$
p_\alpha (y;-\alpha )
 =  \rec{y}\,  F_\alpha(y^{-\alpha }) =
 {\alpha \over y^{\alpha +1}}\,  M_\alpha(y^{-\alpha }) \,,
 \eqno(5.13)$$
 \pni for $1<\alpha<2\,,$
$$ p_\alpha (y;\alpha -2)
 = \rec{y}\,  F_{1/\alpha}(y) =
 {1 \over \alpha }\,  M_{1/\alpha}(y;) \,.\eqno(5.14)$$
Consequently  we can interpret the fundamental	solutions (4.1a) and
(4.1b) in terms of stable $pdf$, so generalizing the arguments
for the standard diffusion  equation
based on (5.1)-(5.7).
\vsp
We easily recognize
that for $0<\nu  <1$
the fundamental solution for the {\bf Signalling problem}  provides
a unilateral extremal stable $pdf$ in (scaled) time with index of
stability $\alpha = \nu  \,, $	which decays according to (4.3b)
with a power law.
\vsp
In fact, from  (4.1b) and (5.11) we note that,
putting $y=r^{-1/\nu  }= 
\tau  =t \,({\sqrt{a}/ x})^{1/\nu  } >0 \,, $
$$ (x/\sqrt{a})^{1/\nu }\, \Gs(x,t;\nu	) =
p_\nu  (\tau;-\nu ) \,. \eqno(5.15)
$$
This property has been	noted also by Kreiss and Pipkin (1986)
based on (3.8) and  on Feller's   result,
$\, p_\alpha  (t; -\alpha  )  \div {\rm exp} (-s^\alpha) $  for
$0<\alpha <1\,. $
\vsp
As far as the {\bf Cauchy problem}
is concerned, we note that the corresponding fundamental
solution provides  a symmetrical $pdf$ in (scaled) distance
with two 
branches, for $x>0$ and $x<0\,, $
obtained one from the other by reflection.
For large $|x|$ each branch exhibits an exponential decay according to
(4.3) and, only for $1/2\le \nu  <1\,, $
it is the corresponding branch of an extremal  stable $pdf$
with index of stability $\alpha =1/\nu	\,. $
In fact, from (4.1b) and (5.14) we note  that, putting
$y= |r| =\xi = |x|/(\sqrt{a}\, t^\nu)  >0\,, $
$$ 2\nu  \, \sqrt{a}\,t^\nu  \,  \Gc(|x|,t;\nu	)
 =  p_{1/\nu  }(\xi, 1/\nu  -2) \,. \eqno(5.16) $$
This property had to the author's knowledge not been noted:
it properly generalizes the Gaussian property of the $pdf$
found for   $\nu  =1/2\, $ (standard diffusion).
Furthermore, using (3.22),  the moments (of even order)
of $\Gc(x,t;\nu )$ turn out to be for $n=1\,, \,2\,, \, \dots$:
$$ \int_{-\infty}^{+\infty} \!\!\! x^{2n}\, \Gc(x,t;\nu  )  \,dx =
   {\Gamma(2n+1)\over \Gamma(2\nu  n+1)}\, (a t^{2\nu  })^n\,.
\eqno(5.17)$$


\section*{Appendix A. The  time fractional derivatives}
For a sufficiently well-behaved function $f(t)$
($ t\in \RR^+$) and for any positive number $\mu$
we may define the fractional derivative
in two different senses,  that we refer here as to
{\bf Riemann-Liouville derivative}
and {\bf Caputo derivative}, respectively.
Both derivatives are related to the  Riemann-Liouville
fractional integral of order $\mu >0$,
  defined as
$$ \frame{J_t^\mu  \, f(t) :=    \rec{\Gamma(\mu )}\,
\int_0^t (t-\tau)^{\mu-1} \, f(\tau )\, d\tau\,.}
\eqno(A.1)  $$
We note the convention $J_t^0 = I$ (Identity)
and the semigroup property for any $\mu, \nu >0$,
$$ J_t^\mu \, J_t^\nu = \,
   J_t^\nu  \, J_t^\mu =J_t^{\mu +\nu} \,. \eqno(A.2)$$
   \vsp
The fractional derivative of order $\mu >0$ in the 
{\bf Riemann-Liouville} sense  is defined as the operator
$ D_t^\mu$ which is the
left inverse of
the Riemann-Liouville integral of order $\mu $
(in analogy with the ordinary derivative), that is
$$ D_t^\mu \, J_t^\mu  = I\,, \; \mu >0\,. \eqno(A.3) $$
If $m$ denotes the positive integer
such that $m-1 <\mu  \le m\,,$  we recognize from Eqs. (A.2) and (A.3)
$$D_t^\mu  \,f(t) :=  \, D_t^m\, J_t^{m-\mu}  \,f(t)\,,\eqno(A.4) $$
hence for $m-1 <\mu  < m,$
$$\frame{
 D_t^\mu  \,f(t) = 
  {\ds {d^m\over dt^m}}\lt[
  {\ds \rec{\Gamma(m-\mu )}\int_0^t
    {f(\tau)\,d\tau  \over (t-\tau )^{\mu  +1-m}} }\rt] \,,}
	\eqno(A.4a)$$
and for $\mu =m$
$$ D_t^\mu  \,f(t =
   {\ds {d^m \over dt^m} f(t)} \,.
\eqno(A.4b)$$
For completion we define $ D_t^0 = I\,. $
\vsp
On the other hand, the fractional derivative of order $\mu >0$ in the
{\bf Caputo} sense  is defined as the operator
$\,*D_t^\mu$  such that
$$   _*D_t^\mu \,f(t) :=  \, _tJ^{m-\mu } \, _tD^m \,f(t)\,,\eqno(A.5)$$
hence
for $  m-1<\mu  <m,$
$$\frame{
    _*D_t^\mu \,f(t) =  
    {\ds \rec{\Gamma(m-\mu )}}\,{\ds\int_0^t
 {\ds {f^{(m)}(\tau)\, d\tau \over (t-\tau )^{\mu  +1-m}}}} \,,}
 \eqno(A.5a)$$
 and for $\mu =m$
$$ _*D_t^m \,f(t) =
   {\ds {d^m \over dt^m} f(t)} \,,.
\eqno(A.5b) $$
This definition 
requires for non-integer $\mu$ the absolute integrability of the  derivative of order $m$.
Whenever we use the operator   $\,_* D_t^\mu$   we (tacitly) assume that
     this condition is met.
\vsp
 We  easily recognize that in general
the two fractional derivative differ for non integer orders
 unless   the function  $f(t)$ along with its first $m-1$ derivatives
 vanishes at $t=0^+$.
In fact, assuming that
the passage of the $m$-derivative under
the integral is legitimate,
we have
$$\frame{
    D_t^\mu \, f(t) =
  \,_*D_t^\mu   \, f(t) +
  \sum_{k=0}^{m-1} f^{(k)}(0^+) \,   \frac{t^{k-\mu}}{\Gamma(k-\mu +1)}
    \,,} \eqno(A.6)    $$
 and therefore, recalling the fractional derivative of the power
functions 
$$\frame{ \,_*D_t^\mu  \, f(t) =
   \,D_t^\mu \left( f(t) -
 \sum_{k=0}^{m-1}   f^{(k)} (0^+) \, \frac{t^k }{ k!} \right)
        \,.}\eqno(A.7)  $$
		\vsp
From (A.7) we recognize that the Caputo  fractional derivative
represents a sort of regularization in the time origin for the
 Riemann-Liouville fractional derivative.
 \vsp
We also note that for its existence
all the limiting   values 
$f^{(k)}(0^+):= {\ds \lim_{t\to 0^+} \,D_t^{k}f(t)}$
are required to be finite for $k=0,1,  \dots m-1$.
\vsp
In the special case   $f^{(k)}(0^+)=0$  for $k=0,1,\dots  m-1$,
we recover the identity between the two fractional derivatives. 
\vsp
Furthermore we observe that the semigroup property of the standard derivatives is not generally  valid
for both the fractional derivatives when the order is not integer.  
\vsp
We now explore the most relevant differences between the two
fractional derivatives. 
We first  observe their different behaviour    at the end points of the interval $(m-1,m)$,
namely when the order is any positive integer, as it can be noted from their
definitions (A.4), (A.5). 
For $\mu \to m^-$ both derivatives 
reduce to $\,D_t^m$, as explicitly stated in Eqs. (A.4b), (A.5b), 
 due  to the fact that the operator $\,J_t^0 =\, I$ commutes with $\,D_t^m$.
However, for $\mu \to (m-1)^+$ we have: 
$$ 
\begin{cases}
 {\ds \,D_t^{\mu} f(t)} \to   
 {\ds D_t^m \, J_t^1 \, f(t) =  D_t^{m-1} \,f(t)= f^{(m-1}(t)}\,, \\
{\ds \,_* D_t^{\mu} f(t)} \to 
{\ds J_t^1\,   D_t^m \, f(t) =  f^{(m-1)}(t) - f^{(m-1)}(0^+) } \,.
 \end{cases}
  \eqno(A.8)
$$
As a consequence, roughly speaking, we can say that $\, D_t^{\mu}$ is, with respect to its order $\mu\,, $
 an operator continuous  at any positive integer, 
whereas $\, _*D_t^{\mu}$   is an operator only left-continuous.
\vsp
We point out the major utility
of the Caputo fractional derivative
in treating initial-value problems for physical and engineering
applications where initial conditions are usually expressed in terms of
integer-order derivatives. This can be easily seen
using the Laplace transformation.
For the Caputo derivative of order $\mu$ with $ m-1<\mu  \le m $
we have
$$
\begin{array}{ll}
 \!\! \L &\left\{ \,_* D_t^\mu \,f(t) ;s\right\} \! = \!    s^\mu \,  \widetilde f(s)
   -{\ds \sum_{k=0}^{m-1}    s^{\mu  -1-k}\, f^{(k)}(0^+)} \,,\\
&   f^{(k)}(0^+) := {\ds \lim_{t\to 0^+}\, D_t^{k}f(t)\,.}   
\end{array}
\eqno(A.9)$$
The corresponding rule for the Riemann-Liouville
derivative of order $\mu$   is
$$
\begin{array} {ll}
 \!\! \L &\left\{ \,D_t^\mu  \, f(t);s\right\} \! = \!
     s^\mu \,  \widetilde f(s)
   -{\ds \sum_{k=0}^{m-1} s^{m -1-k}\, g^{(k)}(0^+)}\,,\\
   & g^{(k)}(0^+) := {\ds \lim_{t\to 0^+}\, D_t^{k}g(t)\,,}
\;  {\ds g(t):= \,J_t^{m-\mu}\,f(t) \,.}
\end{array}
  \eqno(A.10)
 $$
 \vsp
 Thus the rule  (A.10) is     more cumbersome to be used
than (A.9) since it requires initial values  concerning
an extra function $g(t)$ related to the given $f(t)$.  
However, when all the limiting   values $f^{(k)}(0^+)$
are finite and the order is not integer, we can prove
by  that all $g^{(k)}(0^+)$
 vanish  so that  
the formula (A.10)  simplifies into
$$ \L \left\{ D_t^\mu  \, f(t);s\right\} =
      s^\mu \,  \widetilde f(s) \,,
	  \;  m-1 <\mu< m\,.
	  \eqno(A.11)$$
In the special case   $f^{(k)}(0^+)=0$  for $k=0,1,  m-1$,
we recover the identity between the two fractional derivatives.
\vsp
We note that {\bf the Laplace transform rule (A.9)}
was the starting point of Caputo, see Caputo (1967), Caputo (1969),
for defining his generalized derivative in the late sixties.
\section*{Appendix B. The  plots of the M-Wright function}
To gain more  insight
   of the effect of the parameter $\nu$ on the behaviour of the $M_\nu$ Wright function close 
   to and far from the origin,
     we will adopt both linear and logarithmic scale for the ordinates.
	 In Figs. B.1 and B.2   
we compare the plots of the  $M_\nu (|x|)$-Wright functions
in  $|x| \le 5$ for some rational values  in the ranges  $\nu \in [0,1/2]$
and $\nu \in [1/2, 1]$, respectively.
Thus in Fig. B.1 
 we see the transition from $\exp (-|x|)$ for $\nu=0$
to $1/\sqrt{\pi}\, \exp (-x^2)$ for $\nu=1/2$, whereas
in Fig. B.2 we 
see the transition from   $1/\sqrt{\pi}\, \exp (-x^2)$  for $\nu=1/2$
to the delta functions $\delta(x\pm 1)$ for $\nu=1$.
\begin{center}
 \includegraphics[width=.70\textwidth]{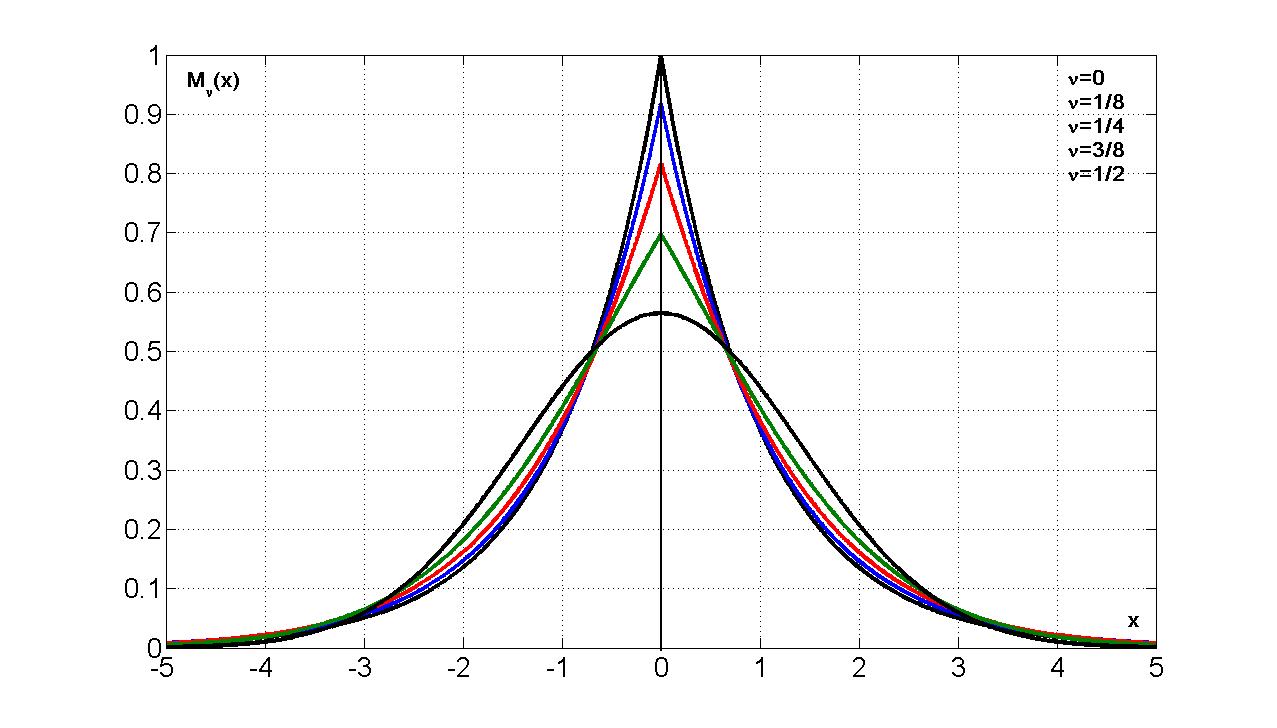}
 \includegraphics[width=.70\textwidth]{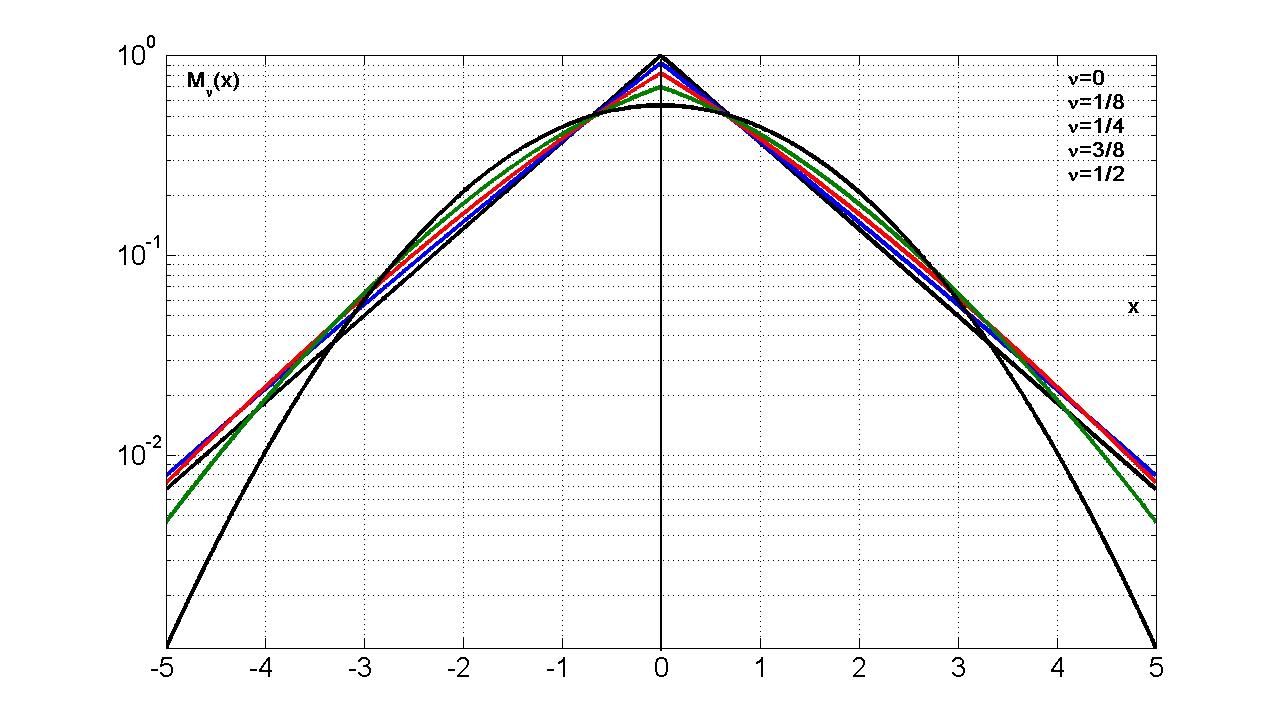}
\end{center}
 \vskip -0.6truecm
\centerline{{\bf Fig. B1} - Plots  of 
 $M_\nu (|x|)$ with $\nu=0, 1/8, 1/4, 3/8, 1/2$ 
 for $ |x| \le 5$;}
\centerline{top: linear scale, bottom: logarithmic scale.}
\newpage
\begin{center}
 \includegraphics[width=.70\textwidth]{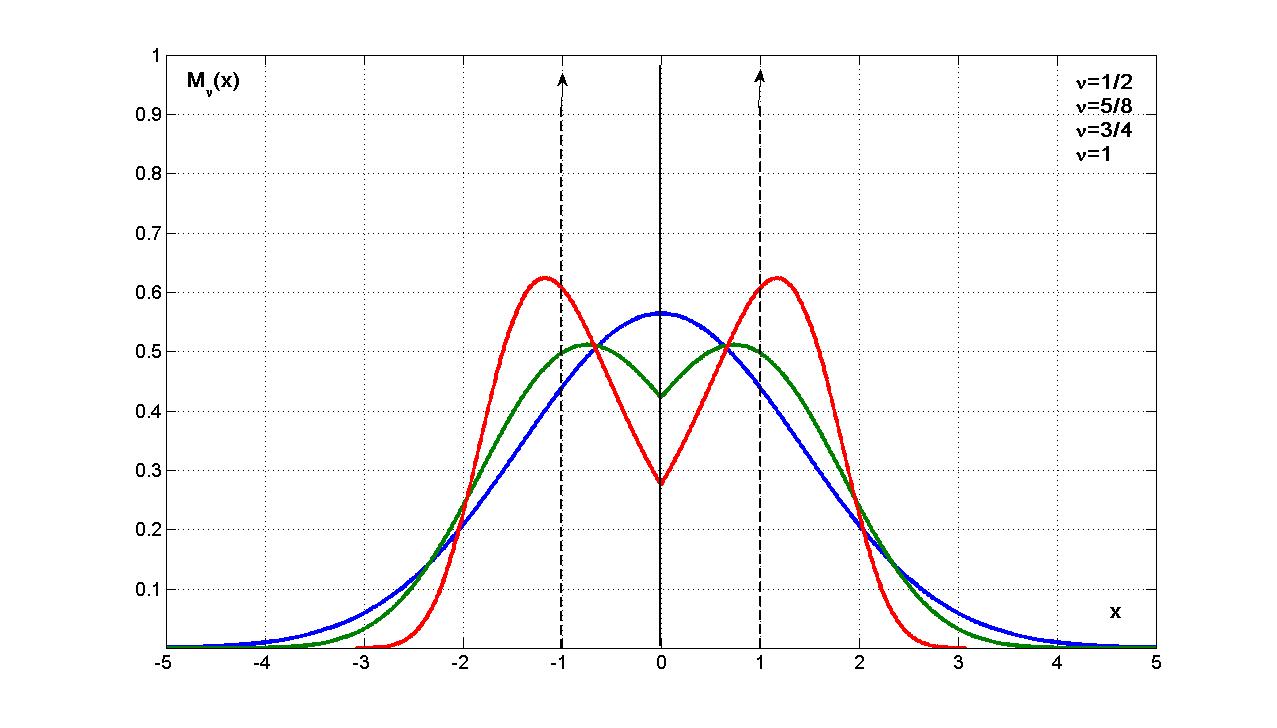}
 \includegraphics[width=.70\textwidth]{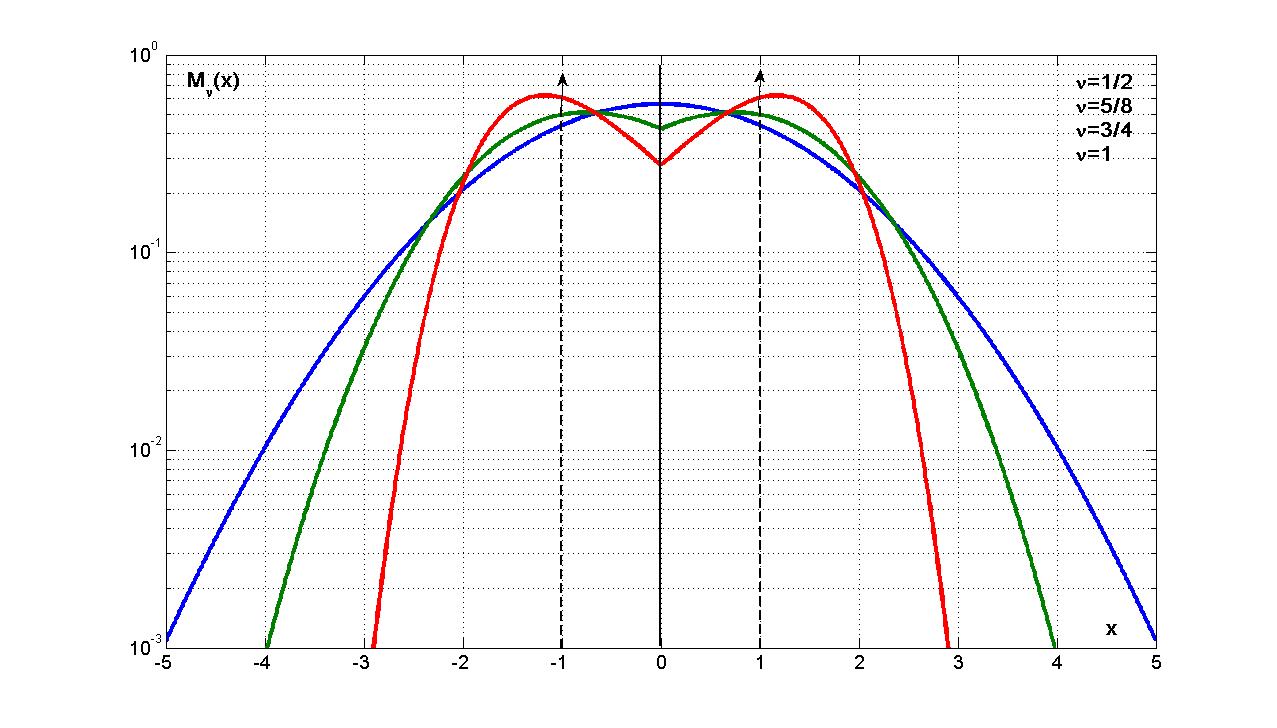}
\end{center}
 \centerline{{\bf Fig. B2} - Plots  of  
 $M_\nu (|x|)$  with $\nu=1/2\,,\, 5/8\,,\, 3/4\,, \,1$ 
 for $ |x| \le 5$:}
 \centerline{top: linear scale; bottom: logarithmic scale)}
  \vsp	 
In plotting $M_\nu (|x|)$ at fixed $\nu $ for sufficiently large $|x|$
the asymptotic representation (3.20) is  useful
 since, as $|x|$ increases,
the numerical convergence of the series in (3.17)
becomes poor and poor up to being completely inefficient.
\vsp
  However, as $\nu \to 1^-$,
the plotting remains a very difficult task because
of the high peaks arising around $x = \pm 1$.
\vsp
In Fig. B.3   we consider the cases $\nu =1 -\epsilon$:  (a) $\epsilon =0.01\,,$ (b) $\epsilon =0.001\,.$
Here the plots are  obtained by the method of Kreis \& Pipkin [28] (continuous line), 
by adding 100 terms-series (dashed line)  and
by the standard saddle-point method (dashed-dotted line).
\begin{center}
 \includegraphics[width=.95\textwidth]{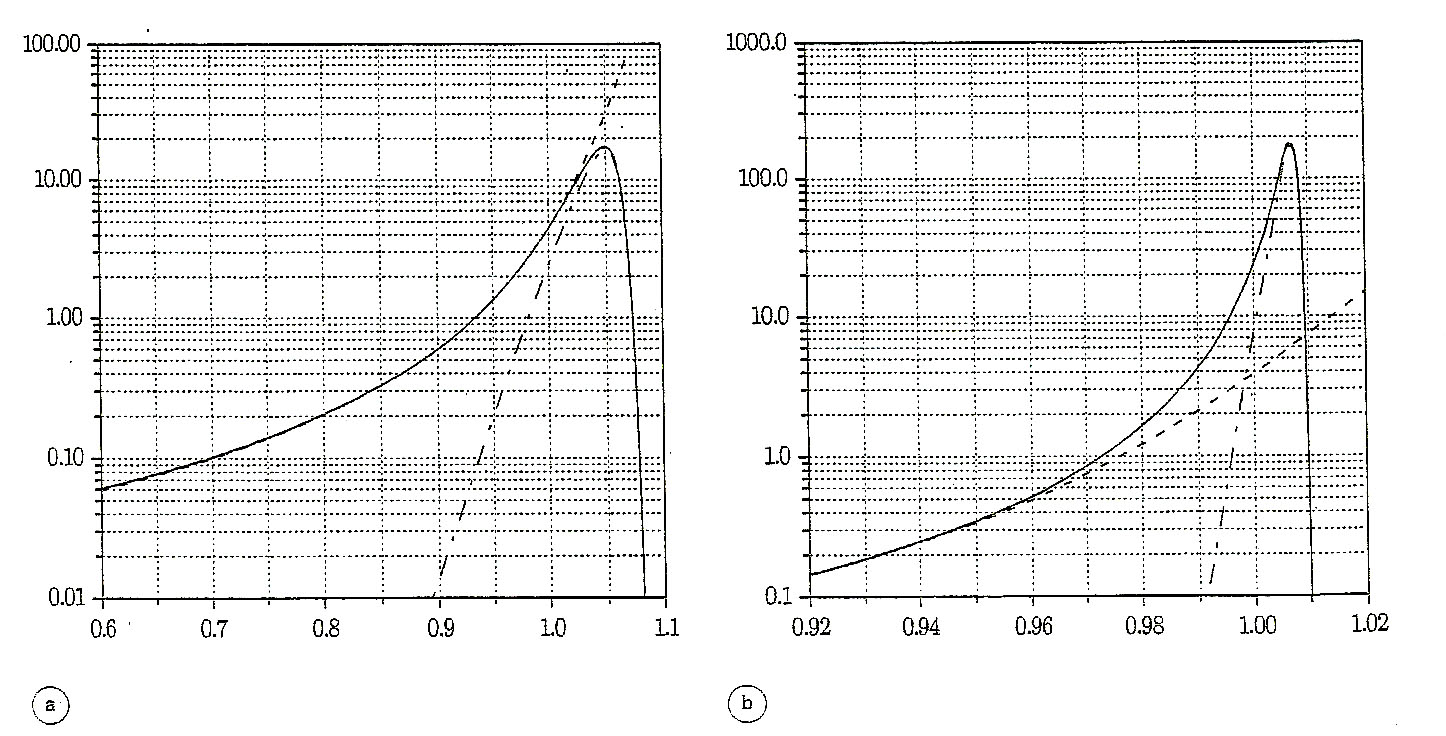}
\end{center}
\centerline{{\bf Fig. B3} -  Plots of $M_\nu(x)$
with $\nu =1-\epsilon$ around the maximum $x \approx 1$;}
\centerline{left: (a) $\epsilon =0.01\,,$ right: (b) $\epsilon =0.001$.}
\vsp



\begin{thebibliography}{99}  
\small





\bibitem{1} Buchen, P.W. and  Mainardi, F. (1975):
    Asymptotic expansions for transient viscoelastic waves,
    {\it  Journal de M\'ecanique}  {\bf 14}, 597-608.
\vsp
\bibitem{2}  Caputo, M. (1966):
  Linear models of dissipation whose Q is almost frequency independent,
  {\it Annali di Geofisica} {\bf 19}, 383-393.
\vsp
\bibitem{3} Caputo, M. (1967):
  Linear models of dissipation whose Q is almost frequency independent,
  Part II.,
  {\it Geophys. J. R. Astr. Soc.} {\bf 13}, 529-539.
  [Reprinted in: {\it Fractional Calculus and Applied Analysis} {\bf 11} (2008), 
  No 1, 3-14.]
\vsp
\bibitem{4} Caputo, M.	(1969): {\it Elasticit\`a e Dissipazione}
  (Zanichelli Bologna). [in Italian]
\vsp
\bibitem{5}  Caputo, M. and   Mainardi, F. (1971a):
A new dissipation model based on memory mechanism, 
{\it Pure and Applied Geophysics (Pageoph)} {\bf 91},  134-147.
[Reprinted in: 
{\it Fractional Calculus and Applied Analysis} {\bf 10} (2007), No 3,  309-324.]
\vsp
\bibitem{6} Caputo, M. and  Mainardi, F. (1971b):
Linear models of dissipation in anelastic solids,
 {\it Rivista del Nuovo Cimento} (Ser II) {\bf 1}, 161-198.
\vsp
\bibitem{7}  Caputo, M. (1973):
  Elasticity with dissipation represented by a simple memory mechanism,
  {\it Atti Accad. Naz. Lincei, Rend. Classe Scienze}  (Ser. 8),
  {\bf 55},  467-470.
 \vsp
\bibitem{8}  Caputo, M. (1976):
  Vibrations of an infinite plate  with a frequency
  independent $Q\,, $
 {\it J. Acoust. Soc. Am.} {\bf 60}, 634-639.
\vsp
\bibitem{9}  Caputo, M. (1979):
  A model for the fatigue in elastic materials with frequency
  independent $Q\,, $
 {\it J. Acoust. Soc. Am.} {\bf 66}, 176-179.
\vsp
\bibitem{10} Caputo, M. (1996):
  The Green function of the diffusion in porous media with memory,
  {\it Rend. Fis. Acc. Lincei} (Ser. 9) {\bf 7}, 243-250.
\vsp
\bibitem{11}  Chin, R.C.Y. (1980):
    Wave propagation in viscoelastic media,
  in {\it Physics of the Earth's Interior},
  edited by A. Dziewonski and E. Boschi
  (North-Holland, Amsterdam), pp. 213-246.
 [E. Fermi Int. School, Course 78]
\vsp
\bibitem{12} Christensen, R.M. (1982): {\it Theory of Viscoelasticity}
(Academic Press, New York).  [1-st ed. (1972)]
\vsp
\bibitem{13}
Dzherbashyan, M.M. and Nersesyan, A.B. (1968):
Fractional derivatives and the Cauchy problem for differential
equations of fractional order.
{\it Izv. Acad. Nauk Armjanskvy SSR, Matematika} {\bf 3}, 3--29.
[In Russian]
\vsp
\bibitem{14} Engler, H. (1997):
Similarity solutions for a class of hyperbolic integro-differential equations,
{\it Differential Integral Equations} {\bf 10}, 815-840.
\vsp
\bibitem{15}  Erd\'elyi, A. {\it Editor} (1955):
  {\it Higher Transcendental Functions},
  Bateman Project (McGraw-Hill, New York), Vol. 3, Ch. 18, pp. 206-227.
\vsp
\bibitem{16} Feller, W. (1971):
{\it An Introduction to Probability Theory and its Applications},
  (Wiley, New York), Vol. II, Ch. 6: pp. 169-176, Ch. 13: pp. 448-454.
[1-st ed. (1966)]
\vsp
\bibitem{17}
 Fujita, Y. (1990a):
  Integro-differential equation which interpolates the heat equation
 and the wave equation, I, II,
  {\it Osaka J. Math.} {\bf 27}, 309-321, 797-804. 
\vsp
\bibitem{18}
 Fujita, Y. (1990b):
  Cauchy problems of fractional order and stable processes,
  {\it Japan J. Appl. Math.} {\bf 7}, 459-476. 
\vsp
\bibitem{19}  Giona, M.	and  Roman, H.E. (1992):
  Fractional diffusion equation for transport phenomena in random
  media,  {\it Physica A} {\bf 185},  82-97.
 \vsp
  \bibitem{20} Gonsovskii, V.L. and Rossikhin, Yu.A. (1973):
  Stress waves in a viscoelastic medium with a singular hereditary kernel,
  {\it Zhurnal Prikladnoi Mekhaniki Tekhnicheskoi Fiziki}
 {\bf  4},  184-186.
[Translated from the Russian by Plenum Publishing Corporation,  New York (1975)]
  \vsp
  \bibitem{21}  Gorenflo, R., Iskenderov, A. and Luchko, Yu.  (2000):
 Mapping between solutions of fractional diffusion-wave equations,
 {\it Fractional Calculus and Applied Analysis} {\bf 3}, 75-86.
\vsp
\bibitem{22} Gorenflo, R. and Mainardi, F. (1997):
 Fractional calculus: integral and differential
 equations of fractional order, in
 {\it Fractals and Fractional Calculus in Continuum Mechanics}, edited by
 A. Carpinteri and F. Mainardi (Springer Verlag, Wien), 223-276.
\vsp
\bibitem{23}  Graffi, D. (1982):
  Mathematical models and waves in linear viscoelasticity,
  in {\it Wave Propagation in Viscoelastic Media}, edited by F. Mainardi
  (Pitman, London), pp. 1-27.  [Res. Notes in Maths, Vol. 52]
 \vsp
\bibitem{24} Hunter, S.C. (1960): Viscoelastic Waves, in
 {\it Progress in Solid Mechanics}, edited by
  I. Sneddon and R. Hill
  (North-Holland, Amsterdam),  Vol 1, pp. 3-60.
\vsp
\bibitem{25} Kilbas, A.A., Srivastava, H.M.  and  Trujillo, J.J. (2006):
 {\it Theory and Applications of Fractional Differential Equations},
 (Elsevier, Amsterdam). 
[North-Holland Mathematics Studies No 204]
 \vsp
\bibitem{26} Kochubei, A.N. (1990):
  Fractional-order diffusion,
  {\it Differential Equations}  {\bf  26},  485-492.
[English translation from the Russian Journal
{\it Differenttsial'nye Uravneniya}]  
\vsp
\bibitem{27} Kolsky, H. (1956): The propagation of stress pulses in viscoelastic
 solids, {\it Phil. Mag.} (Ser 8) {\bf 2}, 693-710.
\vsp
\bibitem{28} Kreis, A. and Pipkin, A.C. (1986):
  Viscoelastic pulse propagation and stable probability distributions,
  {\it Quart. Appl. Math.} {\bf 44}, 353-360.
\vsp
\bibitem{29} Mainardi,  F. (1994):
 On the initial value problem for the fractional
  diffusion-wave equation, in
  {\it Waves and Stability in Continuous Media}
  edited by S. Rionero and T. Ruggeri,
  (World Scientific, Singapore), pp. 246-251.
\vsp
\bibitem{30} Mainardi,  F. (1995):
  Fractional diffusive waves  in viscoelastic solids
  in  {\it IUTAM Symposium - Nonlinear Waves in Solids},
 edited by J. L. Wegner  and F. R. Norwood (ASME/AMR, Fairfield NJ),
  pp. 93-97.
 [Abstract in {\it Appl. Mech. Rev.} {\bf 46} (1993), 549]
\vsp
\bibitem{31} Mainardi, F. (1996a):
 Fractional relaxation-oscillation and fractional
  diffusion-wave phenomena,
  {\it Chaos, Solitons \& Fractals} {\bf 7}, 1461-1477.
\vsp
\bibitem{32} Mainardi, F. (1996b):
  The fundamental solutions for the fractional diffusion-wave
   equation,
   {\it Applied Mathematics Letters} {\bf 9}, No 6, 23-28.
\vsp
\bibitem{33}    Mainardi, F. (1997):
 Fractional calculus;
  some basic problems in continuum and statistical mechanics,
  in {\it Fractals and Fractional Calculus in Continuum Mechanics},
edited by. A. Carpinteri and F. Mai\-nardi
 (Springer-Verlag, Wien), 291-348.
\vsp
\bibitem{34}  Mainardi, F. (2002a) : 
Linear viscoelasticity,  Chapter 4 
in:  A. Guran, A. Bostr\"om, O. Leroy and G. Maze (Editors),
 {\it Acoustic Interactions with Submerged Elastic Structures, Part IV:
 Nondestructive Testing, Acoustic Wave Propagation and Scattering},
  (World Scientific, Singapore), pp. 97-126. 
  [Vol. 5 on the Series B on Stability, Vibration and Control of Systems]
\vsp
\bibitem{35}      Mainardi, F. (2002b): 
 Transient waves in  linear viscoelastic media,  Chapter 5
 in:  A. Guran, A. Bostr\"om, O. Leroy and G. Maze (Editors),
 { \it Acoustic Interactions with Submerged Elastic Structures, Part IV:
 Nondestructive Testing, Acoustic Wave Propagation and Scattering},
  (World Scientific, Singapore),  pp. 127-161.
\vsp
\bibitem{36} Mainardi, F. (2008):
   {\it Fractional Calculus and Waves in Linear Viscoelasticity}
   (Imperial College Press, London), in preparation.
\vsp
\bibitem{37} Mainardi, F. and  Gorenflo, R. (2007):
Time-fractional derivatives in relaxation processes: a tutorial survey,  
{\it Fractional Calculus and Applied Analysis} {\bf 10}, 269-308. 
[E-print http://arxiv.org/abs/0801.4914]    
\vsp
\bibitem{38}  Mainardi, F. Luchko, Yu. and  Pagnini, G. (2001):
 The fundamental solution of the space-time fractional diffusion equation,
  { \it Fractional Calculus and Applied Analysis}  {\bf  4}, 153-192. 
 [E-print http://arxiv.org/abs/cond-mat/0702419]
\vsp
\bibitem{39} Mainardi, F. and Pagnini, G. (2003):
 The Wright functions as solutions of the time-fractional diffusion   equations,
  {\it Applied Mathematics and Computation}  {\bf 141},  51-66. 
\vsp
\bibitem{40} Mainardi, F. and  Paradisi, P. (2001):
   Fractional diffusive waves,
   {\it Journal of  Computational Acoustics} {\bf 9}, 1417-1436. 
\vsp      
\bibitem{41} Mainardi, F. and Tomirotti, M. (1995): On a special function
  arising  in  the time fractional diffusion-wave equation,
  in {\it Transform Methods and Special Functions, Sofia 1994},
edited by P. Rusev, I. Dimovski and V. Kiryakova,
  (Science Culture Technology, Singapore), pp. 171-183.
\vsp
\bibitem{42} Mainardi, F. and  Tomirotti, M. (1997): Seismic pulse propagation
  with constant $Q$ and stable probability distributions,
  {\it Annali di Geofisica} {\bf 40}, 1311-1328.
\vsp
\bibitem{43} Mainardi, F. and  Turchetti, G. (1975):
  Wave front expansion for transient viscoelastic waves,
 {\it Mech. Res. Comm.}  {\bf 2}, 107-112.
 \vsp
\bibitem{44} Meshkov, S.I. and  Rossikhin, Yu. A. (1970):
Sound wave propagation in a viscoelastic medium whose hereditary
properties are determined by weakly singular kernels, in
{\it Waves in Inelastic Media}, edited by Yu. N. Rabotnov
(Kishniev), pp. 162-172. [in Russian]
\vsp
\bibitem{45}  Metzler, R.,  Gl\"ockle, W.G. and  Nonnenmacher, T.F. (1994):
   Fractional model equation for anomalous diffusion,
   {\it Physica A} {\bf 211},	13-24.
\vsp
\bibitem{46}  Nigmatullin, R.R. (1986):
  The realization of the generalized transfer equation in a medium with
  fractal geometry,
   {\it  Phys. Stat. Sol. B} {\bf 133}, 425-430.
  [English translation from the Russian]
\vsp
\bibitem{47} Pipkin, A.C. (1986): {\it Lectures on Viscoelastic Theory}
  (Springer-Verlag, New York), Ch. 4, pp. 56-76. [1-st ed, 1972]
\vsp
\bibitem{48} Podlubny, I. (1999):
{\it  Fractional Differential Equations }
(Academic Press, San Diego). 
[Mathematics in Science and Engineering, Vol. 198]
\vsp
\bibitem{49} Pr\"usse, J. (1993):
{\it Evolutionary Integral Equations and Applications},
(Birkhauser Verlag, Basel).
\vsp
\bibitem{50}  
 Rabotnov,   Yu. N. (1969):
 {\it Creep Problems in Structural Members},
  North-Holland, Amsterdam. 
 [English translation of  the 1966 Russian edition]
\vsp
\bibitem{51} Rossikhin, Yu. A. and Shitikova, M.V. (1997):
Application of fractional calculus to dynamic problems of
linear and nonlinear hereditary mechanics of solids,
{\it Applied Mechanics Review} {\bf 50}, 15-67.
\vsp
\bibitem{52}  
Rossikhin, Yu.A.  and  Shitikova, M.V. (2007).
Comparative analysis of viscoelastic models involving fractional derivatives of different
orders, 
  {\it  Fractional Calculus and Applied Analysis} {\bf 10} No 2,  111-121.
\vsp
\bibitem{53}  
Rossikhin, Yu.A.  and  Shitikova, M.V. (2010):
Applications of fractional calculus to dynamic problems of solid mechanics:
novel trends and recent results,
{\it Appl. Mech. Review} {\bf 63},  010801/1--52. 
\vsp
 \bibitem{54}  Samko S.G.,  Kilbas, A.A.	and   O.I. Marichev (1993):
   {\it Fractional Integrals and Derivatives, Theory and Applications},
   (Gordon and Breach, Amsterdam).
\bibitem{55} Schneider,	W.R. and  Wyss, W. (1989):
  Fractional diffusion and wave equations,
  {\it J. Math. Phys.}  {\bf 30}, 134-144.
\vsp
\bibitem{56}   
  Scott-Blair,  G.W. (1949):
{\it Survey of General and Applied Rheology},
 Pitman, London. 
 \vsp
\bibitem{57}  Strick, E. (1970):
 A predicted pedestal effect for pulse propagation in
 constant-$Q$ solids,	{\it  Geophysics} {\bf 35},   387-403.
\vsp
\bibitem{58}  Strick, E. (1982):
  Application of linear viscoelasticity to seismic wave propagation,
  in {\it Wave Propagation in Viscoelastic Media}, edited by F. Mainardi
  (Pitman, London), pp. 169-193.  [Res. Notes in Maths, Vol. 52]
\vsp
 \bibitem{59}  Strick, E. and  Mainardi, F. (1982):
 On a general class of constant $Q$ solids,
   {\it Geophys. J. R. Astr. Soc.} {\bf 69}, 415-429	
\vsp
\bibitem{60}   
Uchaikin, V.V. and  Zolotarev, V.M. (1999):
 {\it Chance and Stability. Stable Distributions and their Applications},
 VSP, Utrecht.

\end{thebibliography}
\end{document}